\documentclass[twocolumn,showpacs,preprintnumbers,amsmath,amssymb]{revtex4}

\usepackage{amsthm}

\newtheorem{prop}{Proposition}
\begin{document}

\title{Violations of Bell inequalities as lower bounds on the communication cost 
of non-local correlations}

\author{Stefano Pironio}
\email{spironio@ulb.ac.be}

\affiliation{Service de Physique Th\'eorique, CP 225,
Universit\'e Libre de Bruxelles, 1050 Brussels, Belgium}

\date{February 2003}

\begin{abstract}
To reproduce in a local hidden variables theory correlations that violate Bell 
inequalities, communication must occur between the parties. We show that the 
amount of violation of a Bell inequality imposes a lower bound on the average 
communication needed to produce these correlations. Moreover, for every 
probability distribution there exists an optimal inequality for which the 
degree of violation gives the minimal average communication. As an example, to 
produce using classical resources the correlations that maximally violate the 
CHSH inequality,  $\sqrt{2}-1\simeq 0.4142$ bits of communication are necessary 
and sufficient. For Bell tests performed on two entangled states of dimension 
$d\geq 3$ where each party has the choice between two measurements, our results 
suggest that more communication is needed to simulate outcomes obtained from 
certain non-maximally entangled states than maximally entangled ones. 
\end{abstract}

\maketitle

\section{Introduction}
Characterizing the features of quantum mechanics which differentiate it
from classical theories is an important issue for quantum information theory, as
well as from a fundamental perspective. One such peculiarity is the non-local 
character of quantum mechanics, i.e. the fact that quantum correlations are 
incompatible with local realistic theories. Apart from being one of the most 
intriguing aspects of nature, non-locality is deeply related to several quantum 
information processing tasks \cite{gisin, acin}, and is at the core of quantum 
communication complexity \cite{brassard, zeilinger}.

It was Bell \cite{bell} who first showed that correlations obtained by 
measuring two separated subsystems cannot be explained by a classical realistic 
theory if no communication between the subsystems is allowed. The question which 
then follows is: how much communication is required to reproduce
these correlations ? This is a natural way to quantify the non-local
character of quantum correlations in terms of classical resources. We
will show that the inequalities introduced by Bell forty years
ago not only tell us that some communication is necessary to produce the
correlations but also how much.

The situation we consider is the one encountered in bipartite Bell 
scenarios. Two spatially separated parties, Alice and Bob, receive local 
inputs $x$ and $y$ and subsequently produce outputs $a$ and $b$. We denote by 
$M_A$ the number of possible inputs on Alice's side and by $M_B$ the number of 
inputs on Bob's side and restrict ourselves to the case where a finite number of 
distinct outcomes is associated to each input. The scenario is completely 
characterized by the probabilities $p_{ab|xy}$ that Alice outputs $a$ when given 
$x$ and Bob outputs $b$ when given $y$. We therefore associate to each Bell 
scenario a correlation vector $\mathbf p$ with entries $p_{ab|xy}$. Note that 
these entries satisfy the normalisation constraints 
\begin{equation}\label{norma} 
\sum_{a,b}p_{ab|xy}=1 \qquad \mbox{for} \ \begin{array}{c} x=0,\ldots M_A-1 \\ 
y=0,\ldots M_B-1 \end{array}\ . \end{equation}

In the quantum version of the Bell scenario, Alice and Bob share an entangled 
quantum state on which they perform local measurements. The inputs $x$ and 
$y$ then correspond to the possible settings of their measuring apparatus 
and the outcomes $a$ and $b$ correspond to the results of these measurements.

In the classical version of the Bell scenario, Alice and Bob may use
only classical resources, i.e shared randomness (local hidden variables) and
classical communication, to determine their outcomes $a$ and $b$. 
If the two parties have unrestricted access to shared randomness, the classical
cost of producing the correlations $\mathbf p$ is the minimum amount of
communication they must exchange in a classical protocol to achieve this goal. 
Different measures of this amount of communication are possible:

\textbf{i)} \emph{$C_{w}({\mathbf p})$: Worst case communication}: the maximal 
amount of communication exchanged between Alice and Bob in any particular 
execution of the protocol. See \cite{bct,csirik,bacon,toner}.

\textbf{ii)} \emph{$\bar C({\mathbf p})$: Average communication}: the average
communication exchanged between Alice and Bob, where the average is taken 
over the inputs and the shared randomness. See \cite{maudlin,steiner,methot}.

\textbf{iii)} \emph{$C_{\infty}({\mathbf p})$: Asymptotic communication}: the 
limit $\lim_{n\rightarrow \infty} \bar C({\mathbf p}^{\otimes n})/n$, where 
${\mathbf p}^{\otimes n}$ is the probability distribution obtained when $n$ runs 
of the Bell scenario are carried out in parallel, that is when the 
parties receive $n$ inputs and produce $n$ outputs in one go. See \cite{cgm}.

In each of these definitions the costs are defined with respect to the
optimal protocol that gives the lowest value for each quantity. 

The  asymptotic measure $C_{\infty}$ may be the most appropriate when one is 
concerned with practical applications that make use of the correlations but is 
less preoccupied whether the measurements are performed individually or 
collectively. On the other hand, the first two measures of communication relate 
to protocols where the outcomes are determined after each single pair of inputs 
is chosen. This is in particular the situation encountered in Bell tests. 
These two measures thus more properly count the communication necessary to 
simulate classically non-locality and it could be expected that they are closely 
connected to Bell inequalities. Relations between the worst case situation and 
Bell inequalities were examined in \cite{bacon} where the authors introduced 
new Bell inequalities that are satisfied by correlations that necessitate at 
most 1 bit of communication to be simulated.

In the present paper, we concentrate on the average communication $\bar C$. We 
first point out that the amount by which the probabilities $\mathbf p$ violate 
a Bell inequality imposes a lower bound on $\bar C(\mathbf p)$. This bound is 
simply a bound on the amount of communication needed to classically simulate a 
violation of the inequality. It is a priori unclear that one particular 
manifestation of the non-local content of correlations, the violation of a 
specific Bell inequality, suffices to characterize exactly the communication 
$\bar C(\mathbf p)$ necessary to reproduce the entire set of correlations (all 
the less since in general correlations violate more than one inequality). Yet, 
to each probability distribution $\mathbf p$ is associated an optimal 
inequality such that the bound the violation imposes on $\bar C(\mathbf p)$ is 
saturated, i.e. it gives the minimal average communication needed to reproduce 
these correlations. We then investigate in detail the case of the CHSH 
inequality \cite{chsh}. We show that for two settings and two outcomes Bell 
scenarios, the CHSH inequality is optimal for all quantum correlations. This 
implies in particular that $\sqrt{2}-1\simeq 0.4142$ bits are necessary and 
sufficient on average to reproduce classically the correlations that lead to the 
maximal violation of the inequality. We then apply our approach to the 
CGLMP inequality \cite{cglmp}. We find that for two measurements scenarios more 
communication is needed to reproduce the effect of measuring certain 
non-maximally entangled states of two qutrits than is necessary for maximally 
entangled ones. Our results, combined with those of \cite{adgl}, suggest that 
this is also the case for qu$d$its with $d\geq 3$. Finally we ask whether 
for quantum correlations the optimal inequalities from the communication point 
of view are always facet inequalities. We give an example where this is not 
the case.

This paper is organized as follows. We first describe in section II how the 
average communication $\bar C$ relates to the degree of violation of Bell 
inequalities. We then apply these ideas to the CHSH inequality in section III 
and to the CGLMP inequality in section IV.  In section V, we discuss the 
relations between optimal and facet inequalities.

\vfill

\section{General formalism}
\subsection{Deterministic protocols}
To state our results it is necessary to consider particular classical 
protocols, the deterministic ones which  don't use any kind of randomness. 
These protocols therefore always produce the same pair of outcomes for given 
inputs $x$ and $y$. The entries of the associated correlation vector ${\mathbf 
d}$ are thus of the form $d_{ab|xy}=\delta^a_{\alpha(x,y)}\delta^b_{\beta(x,y)}$ 
where $\alpha(x,y)$ and $\beta(x,y)$ specify Alice's and Bob's outcomes for 
measurements $x$ and $y$. Since there are a finite number of functions 
$\alpha(x,y)$ and $\beta(x,y)$, there are a finite number of different 
deterministic strategies ${\mathbf d}^\lambda$ which we index by $\lambda$. 
Their interest is that any classical protocol can be viewed as a probability 
distribution $\{q_\lambda\}$ of deterministic protocols ${\mathbf d}^\lambda$. 
That is any correlation vector ${\mathbf p}$ can be written as ${\mathbf 
p}=\sum_{\lambda}q_{\lambda}{\mathbf d}^{\lambda}$ where $q_\lambda \geq 0$ and 
$\sum_\lambda q_\lambda=1$. 

Deterministic protocols for which $\alpha$ and $\beta$ depend only 
on the measurements performed locally by each party, i.e. $\alpha=\alpha(x)$,
$\beta=\beta(y)$, are local protocols. No communication at all is required to 
implement them. On the other hand, if $\alpha(x,y)$ or $\beta(x,y)$ depends on 
the input of the other party, some (deterministic) communication $c(x,y)$ 
between the parties is necessary to carry out the protocol.

It will be convenient to group in subsets $\mathcal D_i$ deterministic 
strategies that need the same comunication $c_i$ to be implemented.  Since in 
the present paper we are interested in the average communication $\bar C$, we 
will group the deterministic strategies with respect to the minimal average 
communication needed to implement them, expressed in bits. Indexing strategies 
in $\mathcal D_i$ by $\lambda_i$, we thus have $\bar C({\mathbf 
d}^{\lambda_i})=c_i$ $\forall \lambda_i$. We also arrange the subsets $\mathcal 
D_i$ ($i=0,\ldots N$) in increasing order with respect to their communication 
cost: $c_i<c_{i+1}$. Local deterministic strategies thus belong to $\mathcal 
D_0$ for which $c_0$=0, while the maximum communication cost $c_N$ is associated 
with strategies in $\mathcal D_N$. This occurs when both parties need to send 
the value of their input to the other, so $c_N=\log_2{M_A}+\log_2{M_B}$. We will 
further illustrate this grouping of deterministic strategies in section IV.   

With the above notation, a decomposition of $\mathbf p$ in term of 
deterministic strategies can be written
\begin{equation}\label{cidec}
{\mathbf p}=\sum_i\sum_{\lambda_i}q_{\lambda_i}{\mathbf d}^{\lambda_i} \ . 
\end{equation} 
It then directly follows that the average communication $\bar C(\mathbf p, 
\{q_\lambda\}) $ associated to the protocol (\ref{cidec}) is given by 
\begin{eqnarray} \bar C(\mathbf 
p,\{q_\lambda\})&=&\sum_i\sum_{\lambda_i}q_{\lambda_i} \bar C(\mathbf 
d^{\lambda_i}) \nonumber \\ &=&\sum_i\sum_{\lambda_i}q_{\lambda_i}c_i =\sum_i 
q_ic_i \end{eqnarray} where $q_i=\sum_{\lambda_i}q_{\lambda_i}$ is the 
probability to use a strategy from $\mathcal D_i$. The minimum amount of 
communication $\bar C({\mathbf p})$ necessary to reproduce the correlations 
$\mathbf p$ is the minimum of $\bar C(\mathbf p, \{q_\lambda\})$ over all 
possible decompositions of the form (\ref{cidec}). If there exists a 
decomposition such that $q_0=1$, i.e., if the correlations can be written as a 
convex combination of local deterministic strategies, then $\bar C(\mathbf p)=0$ 
and the correlations are local. If for every decomposition $q_0<1$, the 
correlations are non-local and they violate a Bell inequality.

\subsection{Bell inequalities}
A Bell inequality can be viewed as a vector $\mathbf b$ which associates to 
each probability distribution $\mathbf p$ a number $B(\mathbf p)=\mathbf b \cdot \mathbf 
p$. One particular number is the local bound $B_0=\max_{\lambda_0}\{{\mathbf 
b}\cdot{\mathbf d^{\lambda_0}}\}$. By convexity, every local probability 
distribution $\boldsymbol{\mathit l}=\sum_{\lambda_0}q_{\lambda_0}{\mathbf 
d}^{\lambda_0}$ satisfies the inequality $B({\boldsymbol{\mathit l}})\leq B_0$. 
Correlations $\mathbf p$ that violate it, $B({\mathbf p})>B_0$, are therefore 
non-local. To extract more information from $B(\mathbf p)$ than a simple 
detection of non-locality it is necessary to consider not only the upper bound 
$B_0$ the inequality takes on the local subset $\mathcal D_0$, but also on all 
the other subsets $\mathcal D_i$:  
\begin{equation}\label{bi} 
B_i=\max_{\lambda_i} \{{\mathbf b}\cdot{\mathbf d^{\lambda_i}}\} \ . 
\end{equation}
Given this extra knowledge, a constraint on the decomposition (\ref{cidec}) can 
be deduced from the amount by which $\mathbf p$ violates the Bell inequality. 
This turns into a bound on $\bar C(\mathbf p)$ which is the basis of the present 
paper.

\subsection{Main results}
\begin{prop}
For every inequality $\mathbf b$ and 
probability distribution $\mathbf p$, the following bound holds:
\begin{equation} \label{bound}
\bar C({\mathbf p})\geq \frac{B(\mathbf p)-B_0}{B_{j^*}-B_0}c_{j^*} 
\end{equation}
where $j*$ is the index such that $(B_{j^*}-B_0)/c_{j^*}=\max_{j\neq 0} 
\{(B_{j}-B_0)/c_{j}\}$.
\end{prop}

\begin{proof}
From (\ref{cidec}) and (\ref{bi}), we deduce $B(\mathbf p)=\mathbf 
b \cdot \mathbf p=\sum_i\sum_{\lambda_i}q_{\lambda_i}\mathbf b \cdot \mathbf 
d^{\lambda_i}\leq \sum_iq_iB_i$. Since $\sum_iq_i=1$ we find
\begin{equation}
B(\mathbf p)-B_0\leq \sum_{i\neq 0}q_i(B_i-B_0)
\end{equation}
or
\begin{equation}\label{bound2}
q_{j^*}\geq \frac{B(\mathbf p)-B_0}{B_{j^*}-B_0}-\sum_{i\neq 0,j^*} q_i \frac{B_i-B_0}{B_{j^*}-B_0}
\end{equation}
We thus obtain
\begin{eqnarray}
\bar C(\mathbf p)&=&\sum_i q_i c_i \nonumber \\
&\geq &\frac{B(\mathbf p)-B_0}{B_{j^*}-B_0}c_{j^*}+\sum_{i\neq 0,j^*} 
q_i\left(c_i-\frac{B_i-B_0}{B_{j^*}-B_0}c_{j^*}\right) \nonumber \\ &\geq & 
\frac{B(\mathbf p)-B_0}{B_{j^*}-B_0}c_{j^*} \end{eqnarray}
where in the last line we used $(B_{j^*}-B_0)/c_{j^*}\geq (B_i-B_0)/c_i$ which 
follows from the definition of $j^*$. \end{proof}

The bound (\ref{bound}) the inequality $\mathbf b$ imposes on the average 
communication $\bar C(\mathbf p)$ is proportional to the degree of violation 
$B(\mathbf p)$ times a normalisation factor $\frac{c_{j^*}}{B_{j^*}-B_0}$ 
expressed in units of ``communication per amount of violation''. This naturally 
suggests to rewrite Bell inequalities in natural units where 
$\frac{c_{j^*}}{B_{j^*}-B_0}=1$ so that (\ref{bound}) takes a simpler form:

\begin{prop}
Every Bell inequality $\mathbf b$ can be rewritten in a normalised form 
$\mathbf b'$ such that $B'_i\leq c_i$ $\forall i$. For the normalised inequality 
the bound (\ref{bound}) becomes \begin{equation} \label{bound3} \bar C(\mathbf 
p)\geq B'(\mathbf p) \ . \end{equation}
\end{prop}
\begin{proof}
Define the normalised version of the inequality $\mathbf b$ as
\begin{equation}\label{renorm}
\mathbf b'=\frac{c_{j^*}}{B_{j^*}-B_0}\left({\mathbf b}-\frac{B_0}{M_AM_B}{\mathbf 
1}\right)
\end{equation}
where $j^*$ is taken as in Proposition 1. Note that $\mathbf 1 \cdot 
\mathbf p=M_A M_B$ since the entries of all correlations vector $\mathbf p$ 
satisfy the normalisation constraints (\ref{norma}). The effect of the term 
$-\frac{B_0}{M_AM_B}{\mathbf 1}$ in (\ref{renorm}) is thus to shift the value 
the inequality takes on an arbitrary vector $\mathbf p$ from $B(\mathbf p)$ to 
$B(\mathbf p)-B_0$. We therefore get $B'_i=\max_{\lambda_i} \{{\mathbf 
b'}\cdot{\mathbf d^{\lambda_i}}\}=\frac{c_{j^*}}{B_{j^*}-B_0}(B_i-B_0)\leq c_i$ 
where the last inequality holds by definition of $j^*$. 

We then immediately deduce (\ref{bound3}) since $B'(\mathbf p)={\mathbf 
b'}\cdot{\mathbf p}=\sum_i\sum_{\lambda_i}q_{\lambda_i}\mathbf b'\cdot \mathbf 
d^{\lambda_i}\leq \sum_iq_iB'_i\leq\sum_iq_ic_i=\bar C(\mathbf p)$. \end{proof}

Assuming Bell inequalities are written in this standard way where $B_i\leq 
c_i$, it follows from (\ref{bound3}) that for a given probability distribution 
$\mathbf p$, the inequality that leads to the strongest bound on $\bar 
C(\mathbf p)$ is the one for which $B(\mathbf p)$ takes the greatest value. In 
fact we have:

\begin{prop}
Let $\mathbf b_*$ be the normalised inequality 
that gives the maximum violation $B_*(\mathbf p)=\max_{\mathbf b}\{B(\mathbf 
p)\}$ for the correlations $\mathbf p$, then \begin{equation} \bar C(\mathbf 
p)=B_*(\mathbf p) \end{equation}
\end{prop}

\begin{proof}
This follows from the duality theorem of linear programming 
\cite{lp}. Indeed $B_*(\mathbf p)$ is the solution to the following linear 
programming problem: 
\begin{eqnarray}\label{lp}
&\max& {\mathbf b}\cdot{\mathbf p} \nonumber \\ 
&\mbox{subject to}& {\mathbf b}\cdot{\mathbf d}^{\lambda_i}\leq c_i \qquad 
\forall \lambda_0,\ldots,\lambda_i,\ldots,\lambda_N 
\end{eqnarray}
for the variable ${\mathbf b}$. The dual of that problem is
\begin{eqnarray}\label{lp2}
&\min& \sum_i \sum_{\lambda_i} c_i q_{\lambda_i}=\sum_i c_i q_i \nonumber
\\
&\mbox{subject to}& \sum_i \sum_{\lambda_i} q_{\lambda_i} {\mathbf
d}^{\lambda_i}={\mathbf p} \nonumber \\
& & q_{\lambda_i}\geq 0 \qquad \forall
\lambda_0,\ldots,\lambda_i,\ldots,\lambda_N
\end{eqnarray}
for the variables $q_{\lambda_i}$.
The solution to the dual problem is $\bar C(\mathbf p)$ since it just amounts to 
search for the optimal decomposition $\{q_{\lambda_i}\}$ of ${\mathbf p}$ which 
leads to the lowest average communication (note that the condition 
$\sum_i\sum_{\lambda_i} q_{\lambda_i}=1$ is in fact already implied by the 
normalisation conditions that $\mathbf d^{\lambda_i}$ and $\mathbf p$ satisfy). Now, 
the duality theorem of linear programming states that if the primal 
(dual) has an optimal solution, then the dual (primal) problem also has an 
optimal solution and moreover the two solutions coincide, i.e. $B_*(\mathbf 
p)=\bar C(\mathbf p)$.
\end{proof}

This last result introduces the concept of an optimal inequality $\mathbf b_*$ 
from the communication point of view for the correlations $\mathbf p$. Indeed 
the bounds (\ref{bound}) and (\ref{bound3}) can be interpreted as bounds on the 
communication necessary to simulate classically a violation of the inequality 
$\mathbf b$ by the amount $B(\mathbf p)$. Of course this is also a bound on the 
average communication $\bar C(\mathbf p)$ necessary to reproduce the entire set 
of correlations $\mathbf p$. In general however, more communication may be 
necessary to carry out the latter task than the former. For the optimal 
inequality $\mathbf b_{*}$, though, the communication is identical in the two 
cases. If we quantify non-locality by the amount of communication needed to 
simulate it classically, a violation of the inequality $\mathbf b_{*}$ by the 
amount $B_*(\mathbf p)$ therefore exhibit the complete non-locality contained in 
the correlations $\mathbf p$.

\subsection{Comparing Bell inequalities}
The bound (\ref{bound}) simply expresses that the most efficient strategy to 
simulate a violation of a Bell inequality uses local deterministic protocols 
(which don't necessitate any communication) and deterministic protocols from 
$\mathcal D_{j^*}$ for which the ratio of violation per communication 
$(B_{j^*}-B_0)/c_{j^*}$ is maximal. Indeed, for that strategy a violation of 
$B(\mathbf p)=(1-q_{j^*})B_0+q_{j^*}B_{j^*}$ implies
\begin{equation}\label{qj}
q_{j^*}=\frac{B(\mathbf p)-B_0}{B_{j^*}-B_0}
\end{equation}
and thus a communication $\bar C=q_{j^*}c_{j^*}=\frac{B(\mathbf 
p)-B_0}{B_{j^*}-B_0}c_{j^*}$ which is nothing more than the right-hand side of 
(\ref{bound}).

The bound (\ref{bound}) can thus be viewed as the minimal communication needed 
to produce a given violation of the inequality $\mathbf b$. This allows us to 
compare the amount of violation of different Bell inequalities, possibly 
corresponding to different Bell scenarios. If the inequalities are normalised so 
that $B_i\leq c_i$, the bound (\ref{bound3}) holds and the comparison is even 
more direct: the greater the violation, the greater the non-locality exhibited by 
the inequality.

This way of weighing Bell inequalities is correct however only if $B(\mathbf 
p)\leq B_{j^*}$. Indeed if this is not the case, the strategy just described no 
more works since in (\ref{qj}) $q_{j^*}>1$. Though the bounds (\ref{bound}) and 
(\ref{bound3}) are still valid, it is then in principle possible to infer 
strongest bounds from the violation of the Bell inequality. This should be 
taken into account when comparing Bell inequalities in this way. 

In the remainder of the paper, we will only be concerned with two settings 
Bell scenarios. Note that in that case, $B(\mathbf p)\leq B_{j^*}$ is always 
satisfied for quantum correlations. Indeed the minimal possible communication in 
a (non-local) deterministic protocol is 1 bit and is associated with strategies 
in $\mathcal D_1$. However every quantum correlation of a two settings Bell 
scenario can be reproduced with 1 bit of communication (indeed it suffices for 
one of the parties to send her input to the other so that they are able to 
simulate classically the quantum correlations). It therefore follows that 
$B(\mathbf p)\leq B_1 \leq B_{j^*}$. 

\subsection{Other measures of communication}
The general arguments we presented in this section remain valid 
independently of the precise way communication is counted and the way 
determinist strategies are accordingly partitioned. Depending on the physical 
quantity one is interested in, different measures for the communication 
cost $c_i$ are thus possible. For example to obtain bounds on the average 
communication needed to reproduce quantum correlations in classical protocols 
that use only 1-way communication, the cost of deterministic strategies using 
2-way communication would be taken to be $c=\infty$. Our results therefore apply 
to all averaged-type measures of communication. 

Note that one can also count the communication using Shanon's entropy if 
it's assumed that the parties may perform block coding. This is natural for 
instance if the parties perform several run of the protocol at once as in the 
definition of the asymptotic communication $C_{\infty}$. The resulting bound 
however will not be a lower bound on the asymptotic communication $C_{\infty}$. 
This is because for Bell scenarios corresponding to $n$ runs in parallel, there 
are deterministic strategies than can't be written as the product of $n$ 
one-run deterministic strategies. As $n$ increases, there thus exist new ways 
of decomposing the correlations in term of deterministic protocols that can 
possibly result in lower communication per run but which are not taken into 
account in the one-run decomposition (\ref{cidec}).

Finally, note that computing the communication costs associated to 
deterministic strategies is in general a difficult task. It is a particular 
problem of the field of communication complexity for which several techniques 
have been specially developed \cite{nisan}. However in the case of the CHSH and 
the CGLMP inequality, the bound (\ref{bound}) can easily be deduced.

\section{CHSH inequality}
Let us now focus on the simplest inequality, the CHSH inequality \cite{chsh}. 
The CHSH inequality refers to two settings and two outcomes Bell scenarios. The 
value the inequality takes on an arbitrary vector $\mathbf p$ is 
\begin{eqnarray}\label{chsh} 
\lefteqn{B(\mathbf p)=} \nonumber \\ 
&p(a_0=b_0)+p(b_0\neq a_1)+p(a_1=b_1)+p(b_1=a_0) \nonumber \\ 
&-[p(a_0\neq b_0)+p(b_0=a_1)+p(a_1\neq b_1)+p(b_1\neq a_0)]\nonumber \\ 
\end{eqnarray}
where $p(a_x=b_y)=p_{00|xy}+p_{11|xy}$ and $p(a_x\neq b_y)=p_{10|xy}+p_{01|xy}$. 
The local bound of this inequality is $B_0=2$. The maximal violation of the CHSH 
inequality by quantum mechanics is $2\sqrt{2}$ and is obtained by 
performing measurements on Bell states. On the other hand, the maximum value it 
can take for all possible distributions is 4 when the four terms with a plus 
sign are equal to one. 

To derive a bound on $\bar C(\mathbf p)$ from (\ref{chsh}), we need to compute 
$\max_{j\neq 0} \{(B_{j}-B_0)/c_{j}\}$. Note that in a deterministic protocol, 
either the two parties don't communicate at all, or one of the parties start 
speaking to the other. In the latter case, the minimum communication she can 
send is 1 bit. This implies that the minimum possible average communication 
for non-local deterministic strategies is $c_1=1$. The following 
protocol with entries $d_{ab|xy}=\delta^a_{\alpha(x,y)}\delta^b_{\beta(x,y)}$, 
where 
\begin{equation}\begin{array}{cccc}
\alpha(x,y)=0&\multicolumn{3}{l}{\mbox{for}\ x,y=0,1} \\
\beta(0,0)=0&\beta(1,0)=1&\beta(0,1)=0&\beta(1,1)=0
\end{array}\end{equation}
can be implemented 
with 1 bit of communication. Indeed it suffices for Alice to send the value of 
her input to Bob. Moreover, the value $B(\mathbf d)$ it takes on the inequality 
(\ref{chsh}) is the maximum possible $B(\mathbf d)=4$. It thus follows that 
$\max_{j\neq 0} \{(B_{j}-B_0)/c_{j}\}=(4-2)/1=2$, so that for the CHSH 
inequality the bound (\ref{bound}) becomes \begin{equation}\label{boundchsh} 
\bar C(\mathbf p)\geq \frac{1}{2} B(\mathbf p)-1 \ . \end{equation}

This implies for instance that to reproduce the optimal quantum correlations at 
least $\sqrt{2}-1 \simeq 0.4142$ bits of communication are necessary. Note that 
to reproduce all possible von Neumann measurements on a Bell state 1 bit is 
sufficient \cite{toner}.

Is it possible to find a protocol that reproduces these correlations with 
that amount $\bar C(\mathbf p)=\sqrt{2}-1$ of communication ? It turns out 
in fact that the CHSH inequality is optimal, i.e., the bound (\ref{boundchsh}) 
is saturated, for all quantum correlations. Indeed, quantum correlations satisfy 
the no-signalling conditions: 
\begin{eqnarray}\label{nosign} 
\sum_{b}p_{ab|xy}=\sum_{b}p_{ab|xy'} \quad \forall y,y' \nonumber \\ 
\sum_{a}p_{ab|xy}=\sum_{a}p_{ab|x'y} \quad \forall x,x' 
\end{eqnarray}
which express that Alice's marginal probabilities are independent of Bob's input and 
conversely. For correlations that obey these constraints, we have:

\begin{prop}
$\bar C(\mathbf p)=\frac{1}{2}B(\mathbf p)-1$ bits of communication are 
necessary and sufficient to simulate probabilities $\mathbf p$ that violate the 
CHSH inequality (\ref{chsh}) and satisfy the no-signalling condition 
(\ref{nosign}). \end{prop}

\begin{proof}
As the ``necessary'' part follows from the bound (\ref{chsh}), we just have to 
exhibit a classical protocol that reproduces the correlations with that amount 
of communication. 

First note that when the bound (\ref{bound}) is saturated, it follows from 
the proof of Proposition 1 that the optimal protocol uses only strategies from 
$\mathcal D_0$ and $\mathcal D_{j*}$ and moreover in these subsets only 
strategies that attain the maximal values $B_0$ and $B_{j*}$ on the inequality 
$\mathbf b$ (there could be more than one subset $\mathcal D_{j*}$ if they are 
several indexes $j_*$ for which $(B_{j*}-B_0)/c_{j*}$ is maximum). In our case, 
this implies that the optimal protocol must be built from local strategies 
$\mathbf d^{\lambda_0}$ and from $1$-bit strategies $\mathbf d^{\lambda_1}$ such 
that $\mathbf b \cdot \mathbf d^{\lambda_0}=B_0=2$ and $\mathbf b \cdot \mathbf 
d^{\lambda_1}=B_1=4$.

The entries of the vectors $\mathbf p$ corresponding to the Bell scenario 
associated with the CHSH inequality consists of 16 probabilities $p_{ab|xy}$ 
since  $a$, $b$, $x$ and $y$ each take two possible values. Half of these 
probabilities appear with a plus sign in the CHSH expression (\ref{chsh}) and 
half of them with a minus sign. Since entries 
$d_{ab|xy}=\delta^a_{\alpha(x,y)}\delta^b_{\beta(x,y)}$ of deterministic 
strategies are either equal to 0 or 1, for a deterministic strategy $\mathbf d$ 
to satisfy $B(\mathbf d)=2$, it must contribute to (\ref{chsh}) with one $-$ and 
three $+$. For local strategies, which assign local values $\alpha(x)$ and 
$\beta(y)$ to Alice's and Bob's outcomes, this leaves eight possibilities. 
Indeed if we choose one of the eight entries appearing in (\ref{chsh}) with a 
$-$ sign to be equal to one, the requirement that three entries appearing with a 
$+$ sign must also be equal to one, fully determines the functions $\alpha(x)$ 
and $\beta(y)$. The resulting eight possible local strategies $\mathbf 
d^{\lambda_0}$ ($\lambda_0=0,\ldots 7$) are given in Table I. On the other hand, 
for a deterministic strategy to attain $B(\mathbf d)=4$, it must contribute to 
(\ref{chsh}) with four terms weighted by a $+$. The assignment of outcomes of 
1-bit strategies $\mathbf d^{\lambda_1}$ are either of the form $\alpha(x)$, 
$\beta(x,y)$ (when Alice sends her input to Bob), or $\alpha(x,y)$, $\beta(y)$ 
(when it is Bob who sends his input to Alice). For each of the four possible 
functions $\alpha(x)$, the requirement that all the entries of the 
deterministic vector equal to one appear with a $+$ in the CHSH inequality fixes 
the function $\beta(x,y)$ and similarly for the four possible functions 
$\beta(y)$. There are thus eight protocols in $\mathcal D_1$ that attain the 
bound $B_1=4$. These strategies are given in Table II.

Having characterized the deterministic strategies from which the protocol is 
built, it remains to determine the probabilities $q_{\lambda}$ with which these 
strategies are used. These must be chosen so that 
\begin{equation}\label{probchsh} 
p_{ab|xy}=\sum_{\lambda_0=0}^7q_{\lambda_0}d^{\lambda_0}_{ab|xy}+\sum_{\lambda_1 
=0}^7q_{\lambda_1}d^{\lambda_1 }_{ab|xy} \end{equation}
holds for the 16 entries $p_{ab|xy}$. Let's focus first on the 
entries that enter in (\ref{chsh}) with a $-$ sign. For each of these eight 
entries, the only contribution to the right-hand side of (\ref{probchsh}) 
different from zero comes from a local deterministic strategy 
$\mathbf d^{\lambda_0}$. This therefore fixes the value of the corresponding 
probability $q_{\lambda_0}$. For instance $q_{0_0}=p_{00|10}$ or 
$q_{1_0}=p_{10|11}$. 

We now have to determine the value of the probabilities $q_{\lambda_1}$ so that 
the eight entries $p_{ab|xy}$ that enter (\ref{chsh}) with a $+$ sign satisfy 
(\ref{probchsh}). For simplicity let's focus on one of these entries: 
$p_{00|00}$. Using Tables I and II, equation (\ref{probchsh}) becomes
\begin{equation}
p_{00|00}=q_{0_0}+q_{1_0}+q_{4_0}+q_{0_1}+q_{2_1}+q_{4_1}+q_{6_1}
\end{equation}
or
\begin{equation}
q_{0_1}+q_{2_1}+q_{4_1}+q_{6_1}=p_{00|00}-p_{00|10}-p_{10|11}-p_{01|01} 
\end{equation}
where we replaced each of the probabilities $q_{\lambda_0}$ with their value 
previously determined. From (\ref{chsh}) and using the no-signalling 
conditions (\ref{nosign}) and the normalisation conditions (\ref{norma}), it is 
not difficult to see that the left-hand side of this equation is equal to 
$(B(\mathbf p)-2)/4$. The same argument can be carried for all the seven other 
entries that contribute to the CHSH inequality with a $+$ sign, each time 
finding that the sum of four probabilities $q_{\lambda_1}$ equals $(B(\mathbf 
p)-2)/4$. Taking $q^{\lambda_1}=(B(\mathbf p)-2)/16$ for $\lambda_1=0,\ldots,7$ 
one therefore obtains a solution to (\ref{probchsh}).

The communication associated to this protocol is thus $\bar 
C=\sum_{\lambda}q_{\lambda}\bar C(\mathbf 
d^{\lambda})=\sum_{\lambda_1=0}^7q_{\lambda_1}=\frac{1}{2}B(\mathbf p)-1$. 
\end{proof}

\begin{table}[t]\label{d0}
\caption{The eight local deterministic strategies for which $B(\mathbf d)=2$} 
\begin{tabular}{c|*{8}{c}}
& $\mathbf d^{0_0}$ & $\mathbf d^{1_0}$ & $\mathbf d^{2_0}$& 
$\mathbf d^{3_0}$&$\mathbf d^{4_0}$&$\mathbf d^{5_0}$&$\mathbf 
d^{6_0}$&$\mathbf d^{7_0}$\\ \hline 
$d_{00|00}$&1&1&0&0&1&0&0&0\\ 
$d_{10|00}$&0&0&0&0&0&1&0&0\\ 
$d_{01|00}$&0&0&1&0&0&0&0&0\\
$d_{11|00}$&0&0&0&1&0&0&1&1\\
$d_{00|10}$&1&0&0&0&0&0&0&0\\
$d_{10|10}$&0&1&0&0&1&1&0&0\\
$d_{01|10}$&0&0&1&1&0&0&1&0\\
$d_{11|10}$&0&0&0&0&0&0&0&1\\
$d_{00|01}$&1&1&1&0&0&0&0&0\\
$d_{10|01}$&0&0&0&1&0&0&0&0\\
$d_{01|01}$&0&0&0&0&1&0&0&0\\
$d_{11|01}$&0&0&0&0&0&1&1&1\\
$d_{00|11}$&1&0&1&1&0&0&0&0\\
$d_{10|11}$&0&1&0&0&0&0&0&0\\
$d_{01|11}$&0&0&0&0&0&0&1&0\\
$d_{11|11}$&0&0&0&0&1&1&0&1\\
\end{tabular}
\end{table}
\begin{table}[t]\label{d1}
\caption{The eight 1-bit deterministic strategies for which $B(\mathbf 
d)=4$}
\begin{tabular}{c|*{8}{c}}
& $\mathbf d^{0_1}$ & $\mathbf d^{1_1}$ & $\mathbf d^{2_1}$& 
$\mathbf d^{3_1}$&$\mathbf d^{4_1}$&$\mathbf d^{5_1}$&$\mathbf d^{6_1}$&$\mathbf 
d^{7_1}$ \\ \hline 
$d_{00|00}$&1&0&1&0&1&0&1&0\\
$d_{10|00}$&0&0&0&0&0&0&0&0\\
$d_{01|00}$&0&0&0&0&0&0&0&0\\
$d_{11|00}$&0&1&0&1&0&1&0&1\\                            
$d_{00|10}$&0&0&0&0&0&0&0&0\\
$d_{10|10}$&0&0&1&1&1&0&1&0\\
$d_{01|10}$&1&1&0&0&0&1&0&1\\
$d_{11|10}$&0&0&0&0&0&0&0&0\\
$d_{00|01}$&1&0&1&0&1&1&0&0\\
$d_{10|01}$&0&0&0&0&0&0&0&0\\
$d_{01|01}$&0&0&0&0&0&0&0&0\\
$d_{11|01}$&0&1&0&1&0&0&1&1\\
$d_{00|11}$&1&1&0&0&1&1&0&0\\
$d_{10|11}$&0&0&0&0&0&0&0&0\\
$d_{01|11}$&0&0&0&0&0&0&0&0\\
$d_{11|11}$&0&0&1&1&0&0&1&1\\
\end{tabular}
\end{table}

\section{More dimensions: the CGLMP inequality}

The CGLMP inequality \cite{cglmp} generalises the CHSH inequality for 
$d$-dimensional systems. This inequality refers to measurement scenarios where 
Alice's and Bob's local settings take two values $x,y=0,1$ and each measurement 
gives $d$ possible outcomes $a,b=0,\ldots,d-1$. The value the CGLMP inequality 
takes on an arbitrary vector $\mathbf p$ is \begin{eqnarray}\label{cglmp} 
\lefteqn{B^d(\mathbf p)=\sum_{k=1}^{[d/2]-1}\left(1-\frac{2k}{d-1}\right)} 
\nonumber \\ &&\Big(P(a_0=b_0+k)+P(b_0=a_1+k+1)\nonumber \\ 
&&+P(a_1=b_1+k)+P(b_1=a_0+k)\nonumber \\ 
&&-[P(a_0=b_0-k-1)+P(b_0=a_1-k)\nonumber \\ 
&&+P(a_1=b_1-k-1)+P(b_1=a_0-k-1)]\Big) \end{eqnarray} where 
$P(a_x=b_y+k)=\sum_{b=0}^{d-1}p_{(b+k)b|x,y}$ is the probability that Alice and 
Bob results satisfy $a=b+k \mod d$ when measuring $x$ and $y$. As shown in 
\cite{cglmp}, the local bound of the inequality is $B^d_0=2$. 

When $d=2$ we recover the CHSH inequality and in that case the maximal quantum 
violation is $B^2_{ME}\simeq 2.828$. For $d>2$, the 
(conjectured) maximal violations obtained from maximally entangled qu$d$its 
are given in \cite{cglmp}. For qutrits the maximum is $B^3_{ME}\simeq 
2.8729$ and this value increases with $d$. This suggests that the CGLMP 
inequality exhibits stronger non-local correlations for larger $d$. This has 
been made more precise by connecting the violation of the CGLMP inequality to 
the resistance of the correlations to the admixture of noise \cite{cglmp}. It 
has however been argued in \cite{adgl}, that the resistance to noise is not a 
good measure of non-locality. Quite surprisingly it was also found in 
\cite{adgl} that for $d>2$ the strongest violation of the CGLMP inequality is 
obtained using certain non-maximally entangled states. For qutrits, for 
instance, the maximal violation obtained from a non-maximally entangled state is 
$B^3_{NME}\simeq 2.9149$ which is higher than the maximum $B^3_{ME}\simeq 
2.8729$ for the maximally entangled one. Moreover, this discrepancy between 
maximally and non-maximally states grows with the dimension. This raises 
several questions on how one should interpret and compare these manifestations 
of non-locality. 

A natural answer is through the bound (\ref{bound}). The derivation of the 
bound for the CHSH inequality in the previous section can directly be applied 
to the CGLMP inequality. This yields
\begin{equation}\label{boundcglmp} 
\bar C^d(\mathbf 
p)\geq \frac{1}{2} B^d(\mathbf p)-1 \ . 
\end{equation}
This bound is the same for all the inequalities of the family (\ref{cglmp}), 
and the strength of these different inequalities can therefore simply be 
measured by the degree by which they are violated. This confirms the intuition 
that the non-locality displayed by the CGLMP inequality grows with the 
dimension.

On the other hand, the fact that for $d>2$ the CGLMP inequality is maximally 
violated for non-maximally entangled states translates into more severe 
constraints on the average communication necessary to reproduce correlations 
obtained by measuring certain non-maximally entangled states than maximally 
entangled ones. For instance, for qutrits (\ref{boundcglmp}) implies that $\bar 
C_{ME}\geq 0.4365$ while $\bar C_{NME}\geq 0.4575$. It could however be that for 
maximally entangled states the CGLMP inequality is not optimal and that another 
inequality will impose stronger bounds for states that are maximally entangled. 

To verify that assertion, we numerically solved the linear programming problem 
(\ref{lp2}) for the correlations that maximally violate the CGLMP 
inequality both on maximally and non-maximally entangled states for $d\leq 8$. 
There exist many different algorithms for linear programming and the only 
difficulty in solving (\ref{lp2}) is to characterize the sets $\mathcal D_i$ of 
deterministic strategies and their corresponding communication costs $c_i$. A 
deterministic strategy assigns a definite value $\alpha(x,y)$ to Alice's 
outcomes and $\beta(x,y)$ to Bob's outcomes for each of the four possible pair 
of inputs $\{x,y\}$, To simplify the notation we write $\alpha_x(y)=\alpha(x,y)$ 
and $\beta_y(x)=\beta(x,y)$. There are two possibilities for $\alpha_x$: either 
$\alpha_x$ is constant (cst), i.e., $\alpha_x(0)=\alpha_x(1)$, and given input 
$x$ Alice does not need any information from Bob to determine her output; or 
$\alpha_x\neq\mbox{cst}$, that is $\alpha_x(0)\neq\alpha_x(1)$, and Alice's 
outcome depends not only on her local setting $x$ but also on Bob's one. In that 
case Alice needs one bit of information from Bob to output her result. The 
situation is similar for Bob. This leads to four possible sets of deterministic 
strategies:

\textbf{i)} $\mathcal D_0$: the set of local deterministic strategies for which 
$\alpha_x=\mbox{cst}$ and $\beta_y=\mbox{cst}$ for $x=0,1$ and $y=0,1$. These 
don't need any communication to be implemented: $c_0=0$.

\textbf{ii)} $\mathcal D_1$: the strategies where $\alpha_x=\mbox{cst}$ for 
$x=0,1$ and at least one of the $\beta_y\neq\mbox{cst}$. These strategies 
necessitate 1 bit of communication from Alice to Bob. This set also contains the 
reverse strategies which need 1 bit of communication from Bob to Alice. The 
communication cost associated to $\mathcal D_1$ is therefore $c_1=1$.

\textbf{iii)} $\mathcal D_2$: the protocols where $\alpha_x=\mbox{cst}$ for one 
of the two values $x=0$ or $x=1$, $\alpha_{\bar x}\neq\mbox{cst}$ for the other 
value $\bar x$ and at least one of the $\beta_y\neq\mbox{cst}$. These strategies 
can be implemented by Alice sending one bit to Bob, the value of her input, and 
then Bob sending back to Alice the value of his input if Alice's input equals 
$\bar x$. The average communication exchanged is $3/2$ bits so that $c_2=3/2$. 
This set also contains the strategies where Alice's and Bob's positions are 
permuted.

\textbf{iv)} $\mathcal D_3$: $\alpha_x\neq\mbox{cst}$ and 
$\beta_y\neq\mbox{cst}$ for $x=0,1$ and $y=0,1$. To implement these strategies 
both parties need to know the input of the other, so $c_3=2$.

With this assignment of communication costs to deterministic strategies and 
for the correlations considered ($d\leq 8$), it turns out from the results of 
the numerical optimisation (\ref{lp2}) that the CGLMP inequality is optimal, 
i.e. the bound (\ref{boundcglmp}) is saturated. For these particular 
measurements, those that gives rise to the maximal violation of the CGLMP 
inequality, more communication is thus necessary to reproduce results obtained 
on non-maximally entangled states than maximally entangled ones. 

It is nevertheless possible that these measurements are not optimal to 
detect the non-locality of maximally entangled states. We performed 
numerical searches for $d=3$, optimising the two von Neumann measurements the 
parties may choose. We found that the measurements that necessitate the maximal 
communication to be simulated are the ones that maximise the CGLMP inequality. 
These results therefore suggest that two measurement settings on each side do 
not optimally detect non-locality for $d\geq 3$. It is still possible that the 
simulatation of POVMs on maximally entangled states would necessitate further 
communication. However, concurring with \cite{adgl}, we believe that more 
settings per site and a corresponding new Bell inequality are needed 
\footnote{A candidate might be the inequality introduced by 
Bechmann-Pasquinucci and Gisin (quant-ph/0204122) which is maximally violated 
for maximally entangled states.}.

\section{Optimal inequalities and facet inequalities}
The CHSH and the CGLMP inequalities are special inequalities: they are facet 
inequalities. Local correlations $\boldsymbol{\mathit l}$ are convex 
combinations of a finite number of points, the local deterministic strategies: 
$\boldsymbol{\mathit l}=\sum_{\lambda_0}q_{\lambda_0}\mathbf d^{\lambda_0}$. The 
set of local correlations thus forms a convex polytope. Every polytope can be 
characterized either by its vertices (the local deterministic strategies) or by 
its facets which are a finite set of inequalities $\mathbf b^i$ ($i=1,\ldots, 
M)$ \begin{equation} \boldsymbol{\mathit 
l}=\sum_{\lambda_0}q_{\lambda_0}d^{\lambda_0} \quad \Leftrightarrow \quad 
\mathbf b^i \cdot \boldsymbol{\mathit l}\leq B_0^i \qquad i=1,\ldots M
\end{equation}
Facet inequalities thus form a minimal set of inequalities that fully 
characterize the local correlations. They can therefore be viewed as tight 
detectors of non-locality. Complete sets of facet inequalities are known in 
some cases \cite{fine,pitowsky,werner,zuko}. For two settings, two outcomes 
Bell scenarios, the CHSH is the unique (up to symmetries and besides 
trivial inequalities that are always satisfied by quantum correlations) facet 
inequality. It turns out that it is also optimal with regard to the average 
communication $\bar C$ for all quantum correlations. For Bell scenarios 
involving more outcomes, we have seen that the CGLMP inequality is optimal for 
certain correlations.

Is it the case that for quantum correlations, optimal inequalities are always 
facet inequalities ? Consider for instance the following correlations belonging 
to a two settings, three outcomes Bell scenario: Alice and Bob share the 
maximally entangled state of two qutrits $|\psi\rangle=1/\sqrt{3}(|00\rangle 
+|11\rangle +|22\rangle)$. The measurements they perform consist of each 
carrying out the transformation $|i\rangle \rightarrow e^{i\phi(i)}|i\rangle$, 
followed by a Fourier Transform $U_{FT}$ for Alice and $U_{FT*}$ for Bob and 
then a measurement in the computational basis. The settings of their measuring 
apparatus are thus determined by the three phases they use. For Alice's 
setting $x=0$ and $x=1$ the phases are $(0,0,0)$ and $(0,0,\pi/2)$, while for 
Bob's settings $y=0$ and $y=1$ they are $(0,0,\pi/4)$ and $(0,0,-\pi/4)$. This 
results in the probabilities \begin{eqnarray}
p(a_x=b_y)=\left(5+(-1)^{f(x,y)} 2\sqrt{2}\right)/9 \nonumber \\
p(a_x=b_y+1)=\left(2-(-1)^{f(x,y)} \sqrt{2}\right)/9 \nonumber \\
p(a_x=b_y+2)=\left(2-(-1)^{f(x,y)} \sqrt{2}\right)/9
\end{eqnarray}
where $f(x,y)=x(y+1)$. 

These correlations violate the CGLMP inequality by the amount $B^3(\mathbf 
p)=\frac{2}{3}(1+2\sqrt{2})\simeq 2.5523$. On the other hand, consider the 
inequality (\ref{chsh}), which has to be viewed now as a three outcomes 
inequality, i.e. $p(a_x=b_y)=\sum_{k} p_{kk|xy}$ and $p(a_x\neq b_y)=\sum_{k\neq 
l}p_{kl|xy}$ where the sum over $k$ and $l$ runs from 0 to 2.  The above 
correlations violate this straightforward generalisation of the CHSH inequality 
to more outcomes by the amount $B^{3c}(\mathbf p)=\frac{2}{9}(1+8\sqrt{2})\simeq 
2.7364$. Since for both inequalities $\bar C(\mathbf p)\geq \frac{1}{2}B(\mathbf 
p)-1$, the generalised CHSH inequality is stronger than the CGLMP ones for these 
particular correlations. Moreover, numerically solving the linear problem 
(\ref{lp2}) we found $\bar C(\mathbf p)=0.3682$ so that  the bound $\bar 
C(\mathbf p)\geq 0.3682$ implied by the generalised CHSH is saturated, i.e. the 
inequality is optimal.

The generalised CHSH inequality, however, is not a facet inequality. Indeed, 
for an inequality to be a facet, the local deterministic strategies that attain 
the local bound $B_0$ (i.e. the vertices that belong to the facet) must generate 
a space of dimension one less than the dimension of the polytope, since they 
form its boundary. It is shown in \cite{masanes} that the two settings three 
outcomes polytope lies in a hyperplane of dimension 24. For the inequality 
(\ref{chsh}), it is easily checked that there are only 21 local deterministic 
strategies that attain the limit $B_0=2$. They thus generate at best a space of 
dimension 21 which is less than the expected value of 23 for (\ref{chsh}) to be 
a facet. 

Does there exist a facet inequality that imposes the same bound $\bar C(\mathbf 
p)\geq 0.3682$ as the generalised CHSH inequality ? There exist algorithms 
that compute all the facets of a polytope given its vertices. Using both the 
reverse search vertex enumeration algorithm \cite{algo1} and the double 
description method \cite{algo2} we obtained the complete set of facet 
inequalities of the two settings, three outcomes local polytope which consists 
of 1116 inequalities. The correlations described above violate 23 of these 
inequalities.

Note that there are various ways of writing a Bell inequality which are 
equivalent for local and quantum correlations. Indeed local and quantum 
correlations satisfy the normalisation (\ref{norma}) and no-signalling conditions 
(\ref{nosign}) which we express as the constraints 
\begin{equation}\label{constr}
\mathbf g^j \cdot \mathbf p =G^j \qquad j=1,\ldots J
\end{equation}
For probabilities that satisfy these conditions, the inequality $\mathbf b 
\cdot \mathbf p\leq B$ can be rewritten in the equivalent form
\begin{equation}\label{rewrite}
\left(\mu_0 \mathbf b+\sum_j \mu_j \mathbf g^j \cdot\right) \mathbf p\leq\mu_0 
B+\sum_j \mu_j G^j \end{equation}
In particular, with that rewriting, a facet inequality will remain a facet 
inequality and an inequality which is violated by correlations satisfying 
(\ref{constr}) will still be violated. This can be geometrically understood as 
follows. Probabilities that satisfy the constraints (\ref{constr}) lie in a 
hyperplane $\mathcal G$ of dimension less than the total dimension of the space 
$\mathcal P$ of all vectors $\mathbf p$.  An inequality $\mathbf b \cdot \mathbf 
p\leq B$ defines a half-space in $\mathcal P$. The fact that for probabilities 
in $\mathcal G$, Bell inequalities can be written in different equivalent ways 
corresponds to the fact that they are different half-spaces of $\mathcal P$ that 
have the same intersection with the hyperplane $\mathcal G$. It is shown in 
\cite{masanes} that the dimension of the two settings three outcomes polytope 
(the set of all local correlations), is the same as the hyperplane $\mathcal G$ 
defined by the conditions (\ref{constr}) of normalisation and no-signalling. It 
therefore follows that the rewriting (\ref{rewrite}) based on these constraints 
is the unique way to rewrite Bell inequalities in an equivalent form for local 
correlations.

However, for probabilities which do not satisfy all of these constraints, such 
as non-local deterministic strategies $\mathcal D_{i\neq 0}$, the rewritten 
inequalities (\ref{rewrite}) are not equivalent to the original one. They will 
thus lead to different bounds on $\bar C(\mathbf p)$. The strongest bound on 
$\bar C(\mathbf p)$ a facet  inequality $\mathbf b$ will impose on the 
correlations $\mathbf p$ is the solution to the following linear programming 
problem for the variables $\mu_j$: 
\begin{eqnarray}\label{lp3}
&\max& \mu_0 B(\mathbf p)+\sum_j \mu_j G^j \nonumber \\
&\mbox{subject to}& \left(\mu_0 {\mathbf b}+\sum_j \mu_j \mathbf g^j\right) 
\cdot \mathbf d^{\lambda_i} \leq c_i
\end{eqnarray}
We numerically solved this linear problem for the correlations described above 
and each of the 23 facet inequalities they violate. The strongest bound 
obtained was given by the CGLMP inequality and is $\bar C(\mathbf p)\geq 
0.2764$. 

This examples shows that there exist quantum correlations for which 
the strongest bound on $\bar C(\mathbf p)$ deduced from facet inequalities is 
lower than the (optimal) bound given from a non-facet inequality. This is 
contrary to the common view according to which facet inequalities are 
``optimal'' tests of non-locality \cite{masanes}.

\section{Conclusion}
In summary, we have shown that the average communication necessary to 
simulate classically a violation of a Bell inequality is proportional to degree 
of violation of the inequality. Moreover, to each set of correlations is 
associated an optimal inequality for which that communication is also sufficient 
to reproduce the entire set of correlations. The key ingredient was to compare 
the amount of violation of Bell inequalities not only with the maximum value 
they takes on local deterministic strategies, but also on non-local ones that 
necessitate some communication to be implemented. 

Part of the interest of this work is that it gives a physical meaning to the 
degree of violation of Bell inequalities and thus provides an objective way 
to compare violation of different inequalities. It also gives a new way to view 
and understand Bell inequalities that could shed new light on some of their 
aspects. For instance, it was commonly assumed that facet inequalities are 
optimal tests of non-locality because they are tight ``detectors'' of 
non-locality. However if we measure non-locality by the communication needed to 
reproduce it, in certain situations non-facet inequalities are better ``meters'' 
of non-locality than are facet ones. 

This work also provides a tool to characterize and quantify the 
non-locality inherent in quantum correlations. As a result, for instance, for 
two measurements on each side it seems that the correlations that necessitate 
the most communication to be reproduced are obtained on non-maximally entangled 
states rather than on maximally entangled ones for $d>2$. It would be 
interesting to know whether this is still the case for more settings and if not, 
what is the corresponding Bell inequality.

\begin{acknowledgments}
I would like to thank Jonathan Barrett and Serge Massar for stimulating 
discussions and useful comments on this manuscript. I acknowledge financial 
support from the Communaut\'e Fran\c caise de Belgique under grant ARC 
00/05-251, from the IUAP program of the Belgian governement under grant V-18, 
and from the EU under project RESQ (IST-2001-37559).
\end{acknowledgments}


\end{document}